**Title:** What the reproductive number $\mathcal{R}_0$ can and cannot tell us about COVID-19 dynamics

**Authors:**

Clara L. Shaw; Center for Infectious Disease Dynamics, Department of Biology, The Pennsylvania State University, University Park, PA 16802; cls6630@psu.edu

David A. Kennedy; Center for Infectious Disease Dynamics, Department of Biology, The Pennsylvania State University, University Park, PA 16802; dak30@psu.edu

**Corresponding Author:** Clara L. Shaw

Abstract

The reproductive number $\mathcal{R}_0$ (and its value after initial disease emergence $\mathcal{R}$) has long been used to predict the likelihood of pathogen invasion, to gauge the potential severity of an epidemic, and to set policy around interventions. However, often ignored complexities have generated confusion around use of the metric. This is particularly apparent with the emergent pandemic virus SARS-CoV-2, the causative agent of COVID-19. We address some of these misconceptions, namely, how $\mathcal{R}$ changes over time, varies over space, and relates to epidemic size by referencing the mathematical definition of $\mathcal{R}$ and examples from the current pandemic. We hope that a better appreciation of the uses, nuances, and limitations of $\mathcal{R}$ facilitates a better understanding of epidemic spread, epidemic severity, and the effects of interventions in the context of SARS-CoV-2.

Introduction

With the emergence of SARS-CoV-2, the novel coronavirus responsible for COVID-19, much attention has been given to the reproductive number, $\mathcal{R}$, and its initial state, $\mathcal{R}_0$ (Viceconte and Petrosillo, 2020). $\mathcal{R}_0$ is the expected number of infections generated by an infected individual in an otherwise fully susceptible population (Anderson and May, 1991; Diekmann et al., 1990). Under relatively general assumptions, $\mathcal{R}_0$ can be used to determine the probability an emerging disease will cause an epidemic, the final size of an epidemic, and what level of vaccination would be required to achieve herd immunity (Anderson and May, 1991; Delamater et al., 2019; Heffernan et al., 2005; Roberts, 2007). Therefore, when interpreted correctly, and in conjunction with additional relevant information, it can yield valuable insight. However, misinterpretation may lead to faulty conclusions regarding disease dynamics.

The virus, SARS-CoV-2 emerged in Wuhan, China in late 2019 and has since become pandemic causing over 413,000 deaths worldwide by June 10th, 2020 ("Johns Hopkins University & Medicine Coronavirus Resource Center," 2020) in addition to severe economic distress. Policy makers have relied on estimates of $\mathcal{R}_0$ to tailor control measures (e.g. Ferguson et al., 2020), but these estimates vary tremendously within and between populations around the globe (Figure 1). It is important to understand *why* these estimates vary. It is also important to understand how the utility of $\mathcal{R}$ is limited. Here, we derive and explain some of the key nuances of $\mathcal{R}$ and $\mathcal{R}_0$, paying particular attention to insights and limitations with respect to the emerging pathogen SARS-CoV-2.

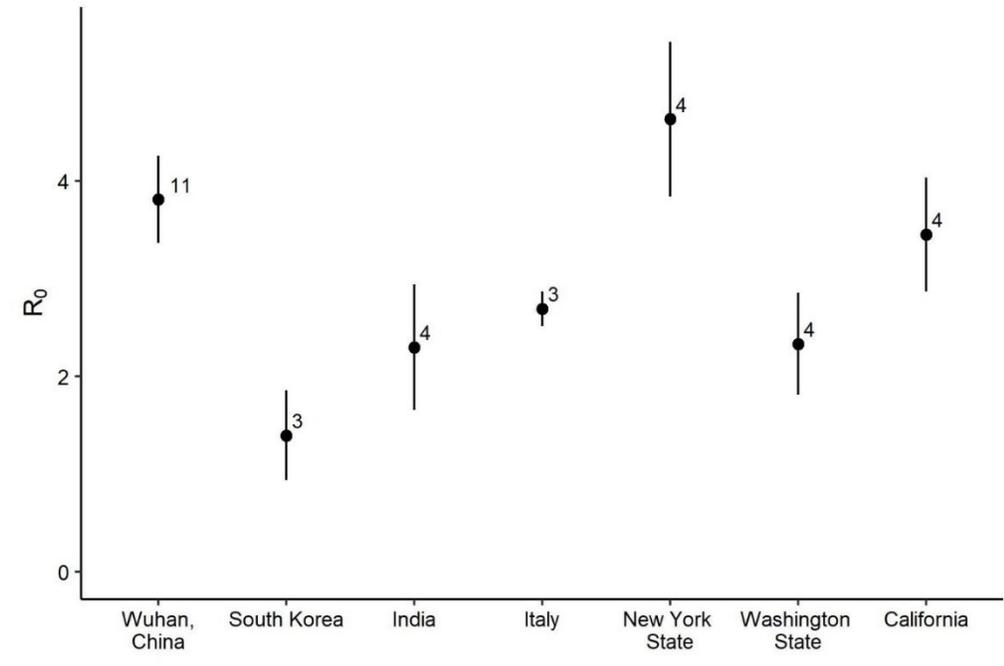

Figure 1. Estimates of the $\mathcal{R}_0$ of SARS-CoV-2 vary substantially between locations affected by the pandemic (Chen et al., 2020; Choi and Ki, 2020; Deb and Majumdar, 2020; Giordano et al., 2020; Johndrow et al., 2020; Korolev, 2020; Lewnard et al., 2020; Q. Li et al., 2020; Tao Liu et al., 2020; Majumder and Mandl, 2020; Mizumoto et al., 2020; Peirlinck et al., 2020; Pitzer et al., 2020; Ranjan, 2020; Read et al., 2020; Riou and Althaus, 2020; Sanche et al., 2020; Senapati et al., 2020; Shim et al., 2020; Singh and Adhikari, 2020; Tang et al., 2020; Wu et al., 2020; Yuan et al., 2020; Zhao et al., 2020). Each point represents the average of a compilation of $\mathcal{R}_0$ estimates from different studies (sample size noted alongside means). Median estimates were used for any studies that provided more than one estimate. Error bars show plus or minus 1 standard error.

Defining $\mathcal{R}$ mathematically

*A general definition*

How many new infections will be caused by a single infected individual? For a directly transmitted pathogen, the answer to this question can be written as:

$$\mathcal{R} = \int_0^\infty k_\tau b_\tau P_\tau \, d\tau \quad (1)$$

Above, $\mathcal{R}$ is the reproductive number, $k_\tau$ is the rate of contacts that an infected individual has with susceptible individuals at time $\tau$ post infection, $b_\tau$ is the probability that a contact at time $\tau$ results in a new infection, and $P_\tau$ is the probability of still being infected at time $\tau$. Notably, we could have combined $b_\tau$ and $P_\tau$ into a single parameter since the probability of infection given contact falls to zero after an individual recovers, but we prefer this more explicit formulation. Equation (1) yields $\mathcal{R}$, the total number of infections one infected individual would generate over the course of their infection. When the population is fully susceptible, as would be expected at the beginning of an outbreak, equation (1) yields $\mathcal{R}_0$. Note that we have neglected to explicitly incorporate individual variation and temporal variation in contact rates, the probability that a contact results in a new infection, and the time to recovery. However, $\mathcal{R}$ and $\mathcal{R}_0$, are intended to be averages, and so this variation is inherently a part of the reproductive number calculation. This variation could be explicitly included with additional subscripts to denote all possible infected hosts, noninfected hosts, times since the epidemic began, and ages of infections. For simplicity, hereafter, we will not use any subscripts for $k$, $b$, and $P$.

*Calculating $\mathcal{R}_0$ using an epidemiological model*

Estimating the individual parameters in equation (1) requires extensive data collection for a specific pathogen, host population, and time. Alternative approaches to estimating $\mathcal{R}_0$ therefore frequently rely on epidemiological models and epidemic data.

Following the lead of Kermack and McKendrick (1927), epidemics have often been modeled as a set of ordinary differential equations. In the simplest Susceptible-Infectious-Recovered (SIR) model, the state variables $S$, $I$, and $R$ are the densities of susceptible, infectious, and recovered individuals in a population. Note that "$R$" here is distinct from the reproductive number "$\mathcal{R}$", but we use both for historical reasons.

$$\frac{dS}{dt} = -\beta SI \quad (2.1)$$

$$\frac{dI}{dt} = \beta SI - \gamma I \quad (2.2)$$

$$\frac{dR}{dt} = \gamma I \quad (2.3)$$

Here, $\beta$ is the transmission coefficient of the pathogen and $\gamma$ is the rate of recovery of infected individuals. Epidemiological data can be used to infer the values of model parameters, including for stochastic or more complex model formulations (for example, Kennedy et al., 2018).

To derive $\mathcal{R}_0$ from these equations, we put the SIR parameters in the context of equation (1). The per infected individual transmission rate, $\beta S$, is equivalent to the contact rate an infected individual has with susceptible individuals, $k$, multiplied by the probability of a new infection resulting from a contact, $b$. Above, the rate of recovery $\gamma$ is constant over time, meaning that the time individuals remain infected is exponentially distributed in this SIR model. The probability of remaining infected, $P$, is thus equal to $e^{-\gamma\tau}$, where $\tau$ is the time since infection. Therefore,

$$\mathcal{R}_0 = \int_0^\infty \beta S e^{-\gamma\tau} d\tau \quad (3.1)$$

Since $S$ changes slowly at the beginning of an epidemic when $I$ is small, we can treat it as a constant with respect to $\tau$. This assumption allows us to analytically solve the integral, which yields

$$\mathcal{R}_0 = \frac{\beta S}{\gamma} \quad (3.2)$$

However, most biological systems are unlikely to conform to the assumptions of this simple SIR model. The model can be modified to better reflect the biology of the system being modeled (for examples, see Keeling and Rohani, 2007) but added complexity can make $\mathcal{R}_0$ more difficult or impossible to solve analytically. In these cases, $\mathcal{R}_0$ can be calculated as the dominant eigenvalue of the next generation matrix (Diekmann et al., 1990) or by other mathematical methods (Heffernan et al., 2005).

*Calculating $\mathcal{R}_0$ without an epidemiological model*

$\mathcal{R}_0$ can also be estimated without an epidemiological model, which can be especially useful if parameter estimates or even an appropriate model structure are not yet known. In principle, one could calculate $\mathcal{R}_0$ by simply counting the cases attributed to infected individuals at or near the beginning of an outbreak. In practice, this method is rarely employed since contact tracing networks are rarely established during the earliest phase of an emerging disease outbreak (but see Pung et al., 2020) and estimates could be inaccurate due to bias towards observing large chains of transmission.

$\mathcal{R}_0$ can also be inferred from the growth rate of cases early in an outbreak. Since the number of susceptible individuals changes slowly during the initial stages of an outbreak, early case growth rates can be approximated by exponential growth: the number of cases $I_t = I_0 e^{rt}$, where $r$ is the epidemic growth rate. If the number of cases $I$ is known for at least two time points, one could calculate the epidemic growth rate $r = \frac{\ln\left(\frac{I_t}{I_0}\right)}{t}$. The relationship between $\mathcal{R}_0$ and $r$ depends on the distribution of the generation interval $T_c$, which is defined as the amount of time between infection of two individuals where the second infection is caused by the first. $T_c$ can be approximated by direct observation or specified by an epidemiological model (Wallinga and Lipsitch 2007). For the model presented in eqs. (2.1)-(2.3) the generation interval is exponentially distributed, and therefore $\mathcal{R}_0 = 1 + rT_c$ (Wallinga and Lipsitch 2007).

No matter the method used to calculate it, limited data or unreliable data early in an epidemic can make it difficult to constrain $\mathcal{R}_0$. The World Health Organization originally estimated the $\mathcal{R}_0$ of SARS-CoV-2 to be between 1.4 and 2.5 (WHO, 2020). More recent estimates of $\mathcal{R}_0$ have varied from 2.2 to 6.47 for the beginning of the Wuhan outbreak (Figure 1). This represents tremendous uncertainty when attempting to use $\mathcal{R}_0$ for public health planning. For example, if we were using these estimates to design a vaccine campaign capable of achieving herd immunity, our vaccination target (calculated as $1 - 1/\mathcal{R}_0$ under assumptions of eq. 2.1-2.3) would be 29% of the population at $\mathcal{R}_0$=1.4 or 85% at $\mathcal{R}_0$=6.47. Even with better estimates of $\mathcal{R}_0$, however, misconceptions around this metric lessen its practical utility.

Misconception 1: $\mathcal{R}_0$ explains future dynamics

As we have explained, $\mathcal{R}_0$ is calculated during the early stages of an epidemic because of its value in determining future infection dynamics (Anderson and May, 1991; Ma and Earn, 2006). But $\mathcal{R}$ changes over time in two important ways that limit its value in understanding future dynamics: first as awareness of infection leads hosts to alter their behavior, and second, as outbreaks progress and new hosts become limiting.

Shifts in behavior that influence contact rates $k$ or the probability of infection given contact $b$ can alter the reproductive number $\mathcal{R}$ over extremely short timescales. For example, as awareness of the SARS-CoV-2 epidemic grew in the United States in March 2020, human mobility ground to a near halt (Gao et al., 2020; Warren and Skillman, 2020), presumably reducing contact rates $k$. Other individual behavioral changes such as increased handwashing and mask wearing (Belot et al., 2020; Goldberg et al., 2020) have likely reduced the probability of transmission given contact $b$ (Liang et al., 2020). Such bottom-up forces combined with top-down government-imposed interventions (e.g. school closures, banned gatherings) reduced $\mathcal{R}$ to below 1 (the threshold for epidemic persistence) by late April in some states (Johndrow et al., 2020; Miller et al., 2020). Similar reductions to $\mathcal{R}$ were documented in China (R.

Li et al., 2020; Tian et al., 2020) and other countries (Ensser et al., 2020; Giordano et al., 2020; Yuan et al., 2020). Indeed, in models of the 1918 influenza pandemic, incorporating a behavioral response to death rates improved model fits (Bootsma and Ferguson, 2007; He et al., 2013). For SARS-CoV-2, behavioral changes may cause $\mathcal{R}$ to fluctuate above and below 1 at different times based on the perceived threat of COVID-19. If that is the case, behavioral fluctuations and intermittent lockdowns may prevent hospital capacity from becoming overwhelmed (Tuite et al., 2020).

While behavioral changes can temporarily reduce $\mathcal{R}$ as described above, more sustainable reductions in $\mathcal{R}$ are typically achieved when susceptible individuals are removed from populations either through naturally acquired immunity or vaccination. However, the impact of removing susceptible individuals is often more complicated than under assumptions of classical SIR models such as eq. 2.1-2.3. When transmission rates are heterogeneous within a population, meaning that some individuals are more likely to contract infection than others, $\mathcal{R}$ declines faster than predicted by eq. 2.1-2.3 (May and Anderson, 1987). This is because those most susceptible (for example, due to high exposure or lower inherent immunity) will become infected earlier in an epidemic, leaving a susceptible population that is on average more resistant (Gomes et al., 2020; Langwig et al., 2017; May and Anderson, 1987). This fact inherently limits the utility of the classical formulation of the "herd immunity" threshold, $1 - 1/\mathcal{R}_0$. Heterogeneity could be included in models to calculate a more realistic herd immunity threshold, but for SARS-CoV-2, data describing heterogeneity in infection risk are still highly uncertain and likely to continue changing through time due to individual or government-mandated responses (Dolbeault and Turinici, 2020). Indeed, current estimates of the fraction of people infected with SARS-CoV-2, in the hardest hit communities (e.g. 15.5% in a small German town exposed to a super spreading event (Streeck et al., 2020) and 19.9% in New York City ("Information on novel coronavirus," 2020)) may be approaching thresholds required for herd immunity calculated under assumptions of extreme heterogeneity conditions (Gomes et al., 2020), but if heterogeneity in transmission is not extreme or if it decreases in the future, herd immunity thresholds will also shift higher.

Misconception 2: The reproductive number is constant over space

The $\mathcal{R}_0$ of many pathogens are often referred to as known values. For example, the $\mathcal{R}_0$ of measles is 12-14, polio is 5-7, and pertussis is 12-17 (Doherty et al., 2016). For SARS-CoV-2, estimates typically range from 2-3 (Ying Liu et al., 2020). However, the parameters ($k$, $b$, and $P$) that make up $\mathcal{R}_0$ can differ substantially from place to place (Figure 1, Delamater et al., 2019). It follows that interventions to reduce $\mathcal{R}$ to less than 1 may need to vary in aggressiveness across locations (Stier et al. 2020).

Since $\mathcal{R}_0$ differs between groups of people, combining multiple groups together to estimate a population-wide $\mathcal{R}_0$ can produce misleading notions of disease spread. For example, though high measles vaccination rates in the United States keep $\mathcal{R}$ below 1 nation-wide, smaller unvaccinated communities still experience serious outbreaks (Leslie et al., 2018). In the current COVID-19 pandemic, disease transmission has so far been much higher in refugee and low income populations compared to non-refugee and high income populations (Chopra and Sobel, 2020; Lau et al., 2020; Ruiz-Euler et al., 2020). Since $\mathcal{R}_0$ is an average, combining communities with high and low transmission may yield an estimate of $\mathcal{R}_0$<1, yet disease may still readily spread (Li et al., 2011). On the other hand, splitting populations may mean missing transmission events that occur between populations, thus underestimating $\mathcal{R}_0$ (Smith et al., 2009).

Awareness of the consequences of how people are grouped can help us interpret $\mathcal{R}$ values. Within groups, behavior associated with work, home, and recreation affects contact rates $k$. As we have discussed above, SIR models often assume that contact rates, and thus $\mathcal{R}_0$, depend on host density. Although, evidence is mixed as to whether larger cities have higher values of $\mathcal{R}_0$ for SARS-CoV-2 (Heroy, 2020; Stier et al., 2020), built environments (e.g. hospitals, airport terminals, factories) do often have high values of $\mathcal{R}_0$ (Dietz et al., 2020). This may partially explain patterns of explosive transmission in venues such as cruise ships and meat packing facilities (Althouse et al., 2020; Dyal et al., 2020; Mizumoto and Chowell, 2020). Household contacts have also been particularly important to SARS-CoV-2 transmission dynamics (Bi et al., 2020). Therefore, differences in $\mathcal{R}_0$ between populations could in part be due to the difference in household sizes between countries and cultures (Gardner et al., 2020; Singh and Adhikari, 2020; Su et al., 2020). Similarly, the probability a new infection results from contact $b$ and probability of remaining infected over time $P$ may vary by population. For SARS-CoV-2, individuals with severe symptoms have 60 times more viral RNA in nasal swabs, which likely increases their ability to transmit the virus and the amount of time they remain infected (Yang Liu et al., 2020). Since older individuals are more likely to develop severe infection (Yang et al., 2020), $\mathcal{R}_0$ is likely to be greater in populations with older individuals, such as in nursing homes (McMichael et al., 2020) or in developed countries (Dowd et al., 2020).

Misconception 3: The reproductive number is enough to tell us how large an epidemic will be

It is tantalizing to imagine that $\mathcal{R}_0$ can be used to predict the extent an outbreak, since it can be calculated during the early stages of an epidemic. Indeed $\mathcal{R}_0$ is related to final epidemic size (Kermack and McKendrick, 1927), but this relationship can be substantially affected by the fraction of the population infected initially and heterogeneity in transmission.

If we rescale population sizes such that the initial susceptible population size $S_0 = 1$, and we assume that population sizes are sufficiently large to neglect demographic stochasticity (Hartfield and Alizon, 2013; Tildesley and Keeling, 2009), then $Z$, the fraction of the population infected during an epidemic, can be determined from the final epidemic size equation, $Z = S_0(1 - e^{-\mathcal{R}_0(Z-I_0)})$. Note that this equation prominently features $I_0$, the fraction of the population infected at the beginning of the outbreak (or at the beginning of an intervention). Efforts to reduce the reproductive number below 1 are understandably a high priority, but when $\mathcal{R}_0$ is close to or less than 1, the final outbreak size is more sensitive to changes in the fraction of infected individuals $I_0$ than it is to changes in $\mathcal{R}_0$ (Figure 2). $\mathcal{R}_0$ is a critical threshold for pathogen invasion, which matters when $I_0$ is small, but which becomes less important as $I_0$ gets larger (Figure 2). Now that SARS-CoV-2 infection rates are already substantial in populations around the globe, $I$ must be considered in addition to $\mathcal{R}$. For example, Pei et al. (2020) estimated that implementing social distancing policies one week earlier could have reduced the cases in the United States by early May, 2020 by 55% (over 700,000 cases) by keeping the number of infected individuals low at the time such policies were implemented.

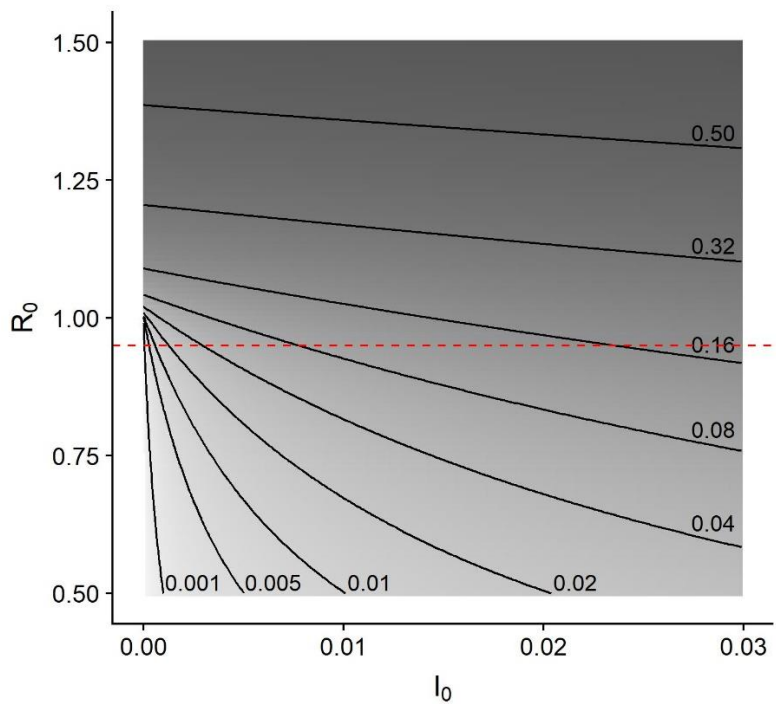

Figure 2. Epidemic size contours and shading show that when $\mathcal{R}_0$ is close to 1, the epidemic is more strongly influenced by a reduction of $I_0$ than by a reduction of $\mathcal{R}_0$. For instance, if $\mathcal{R}_0 = 0.95$ (red dashed line), the epidemic could infect from less than 0.1% to greater than 16% of the population as $I_0$ ranges from 0% to 3% of the population. Epidemic size was calculated using the final size equation, $Z =$

$S_0(1 - e^{-\mathcal{R}_0(Z-I_0)})$, where $S_0=1$. Shading indicates the cube root of epidemic size with lighter colors corresponding to smaller outbreaks.

While a final epidemic size can be calculated using $\mathcal{R}_0$ and $I_0$, the final size equation above does not apply to populations with heterogeneous infection risk (Andreasen, 2011; Ball, 1985; Hébert-Dufresne et al., 2020; Ma and Earn, 2006). Heterogeneity could in principle be incorporated into the final epidemic size equation (Dwyer et al., 2000), but estimating heterogeneity early in an epidemic can be challenging. Moreover, as we describe in misconception 1, heterogeneity in infection risk can change over time as a result of human behavior or interventions, such as for example the shutdowns in response to the COVID-19 epidemic (Dolbeault and Turinici, 2020; Ruiz-Euler et al., 2020). Estimates of how future government restrictions and behavioral changes will alter heterogeneity in infection risk are thus critical for assessing the likely impact of the outbreak (Gomes et al., 2020). Such estimates are also key in determining thresholds for herd immunity and in prioritizing the distribution of interventions such as vaccines when they first become available (Atkinson and Cheyne, 1994; Giambi et al., 2019).

Conclusions

As we have discussed, the reproductive number $\mathcal{R}$ and its initial value $\mathcal{R}_0$ can be used to assess the potential for disease invasion and persistence, to predict the extent of an epidemic, and to infer the impact of interventions and of relaxing control measures. However, the utility of $\mathcal{R}$ and $\mathcal{R}_0$ can easily be overstated. We have focused on three misconceptions that can lead to inaccurate perceptions of disease dynamics. These misconceptions are problematic no matter the complexity of the model or the reliability of the data used to estimate $\mathcal{R}$ because populations vary over space and time and in their changing responses to disease. Considering these nuances when interpreting $\mathcal{R}$ allows for a stronger understanding of the patterns with which SARS-CoV-2 virus has traversed the globe, why it has impacted some populations more than others, and how best to limit future transmission.

Acknowledgments

We thank Amrita Bhattacharya for feedback on earlier versions of the text. This work was supported by startup funds from The Pennsylvania State University. DAK was also partially supported by National Science Foundation grant DEB-1754692. The funders had no role in study design, data collection and analysis, decision to publish, or preparation of the manuscript.